\documentclass[useAMS,usenatbib,usegraphicx]{mn2e}

\title[Microlensing variability in FBQ 0951+2635]{Microlensing variability in FBQ 
0951+2635: short-timescale events or a long-timescale fluctuation?}
\author[V. N. Shalyapin et al.]{V. N. Shalyapin$^{1}$\thanks{E-mail: vshal@ire.kharkov.ua 
(VNS); goicol@unican.es (LJG); koptelova@xray.sai.msu.ru (EK); artamon@sai.msu.ru (BPA); 
alexey@astron.kharkov.ua (AVS); zheleznyak@astron.kharkov.ua (APZ); 
talat77@rambler.ru (TAA); boa@astrin.uzsci.net (OAB); snurit2006@yahoo.com (SNN); 
ullanna@inta.es (AU)}, L. J. Goicoechea$^{2}$\footnotemark[1], 
E. Koptelova$^{3,4}$\footnotemark[1], B. P. Artamonov$^{3}$\footnotemark[1],
\newauthor
A. V. Sergeyev$^{5}$\footnotemark[1], A. P. Zheleznyak$^{6}$\footnotemark[1], 
T. A. Akhunov$^{7,8}$\footnotemark[1], O. A. Burkhonov$^{7}$\footnotemark[1], 
\newauthor
S. N. Nuritdinov$^{7,8}$\footnotemark[1] and A. Ull\'an$^{9}$\footnotemark[1]\\
$^{1}$Institute for Radiophysics and Electronics, National Academy of Sciences of 
Ukraine, 12 Proskura St., 61085 Kharkov, Ukraine\\
$^{2}$Departamento de F\'{\i}sica Moderna, Universidad de Cantabria, Avda. de Los 
Castros s/n, 39005 Santander, Spain\\
$^{3}$Sternberg Astronomical Institute of Moscow University, Universitetski 13, 
119992 Moscow, Russia\\
$^{4}$Graduate Institute of Astronomy of National Central University, Jhongli City, 
Taoyuan County 320, Taiwan\\
$^{5}$Institute of Radio Astronomy, Krasnoznamennaya 4, 61002 Kharkov, Ukraine\\
$^{6}$Institute of Astronomy of Kharkov National University, Sumskaya 35, 61022 
Kharkov, Ukraine\\
$^{7}$Ulugh Beg Astronomical Institute of the Uzbek Academy of Sciences,
Astronomicheskaya 33, 100052 Tashkent, Uzbekistan\\
$^{8}$National University of Uzbekistan, Physics Faculty, 100174 Tashkent, 
Uzbekistan\\
$^{9}$Centro de Astrobiolog\'{\i}a (CSIC-INTA), Ctra de Ajalvir, km 4, 28850 
Torrej\'on de Ardoz, Madrid, Spain}
\begin{document}

%\date{Accepted 1988 December 15. Received 1988 December 14; in original form 1988 
%October 11}

\pagerange{\pageref{firstpage}--\pageref{lastpage}} \pubyear{2009}

\maketitle

\label{firstpage}

\begin{abstract}
We present and analyse new $R$-band frames of the gravitationally lensed double quasar 
FBQ 0951+2635. These images were obtained with the 1.5m AZT-22 Telescope 
at Maidanak (Uzbekistan) in the 2001$-$2006 period. Previous results in the $R$ band 
(1999$-$2001 period) and the new data allow us to discuss the dominant kind of 
microlensing variability in FBQ 0951+2635. The time evolution of the flux ratio $A/B$ 
does not favour the continuous production of short-timescale ($\sim$ months) flares in 
the faintest quasar component B (crossing the central region of the lensing galaxy). 
Instead of a rapid variability scenario, the observations are consistent with the 
existence of a long-timescale fluctuation. The flux ratio shows a bump in the 
2003$-$2004 period and a quasi-flat trend in more recent epochs. Apart from the global
behaviour of $A/B$, we study the intra-year variability over the first semester of 2004, 
which is reasonably well sampled. Short-timescale microlensing is not detected in that 
period. Additional data in the $i$ band (from new $i$-band images taken in 2007 with 
the 2m Liverpool Robotic Telescope at La Palma, Canary Islands) also indicate the 
absence of short-timescale events in 2007.
\end{abstract}


\section{Introduction}

FBQ 0951+2635 was discovered a decade ago by \citet{b12}. This is a double quasar
(consisting of two components A and B) at redshift $z_s$ = 1.246, which is 
gravitationally lensed by an early-type galaxy at $z_l$ = 0.260 \citep{b3}. The main 
lensing galaxy probably belongs to a group of galaxies at similar redshift \citep{b15}. 

\begin{table*}
\centering
\begin{minipage}{150mm}
 \caption{New optical observations of FBQ 0951+2635.}
 \label{MaiLiv}
 \begin{tabular}{@{}lllcccc@{}}
  \hline
  Observatory & Telescope & Camera & Filter
        & Period
        & Nights
        & Exposures/night \\
  \hline
  Maidanak (Uzbekistan) & 1.5m AZT-22 & BroCam & $R$ 
	  	& April 2001 & 1 & 2 (120 + 180 s) \\
  &  &  &   & March 2002 & 3 & 2$-$4 (180/240 s) \\
  &  &  &   & Apr-May 2003 & 2 & 2 (180 s) \\
  &  &  &   & Jan-May 2004 & 14 & 3$-$5 (180/210 s) \\
  &  &  &   & Dec 2004-Apr 2005 & 6 & 4$-$8 (180 s) \\
  &  &  &   & Nov-Dec 2005 & 3 & 2 (300 s) \\
  &  &  &   & Apr-May 2006 & 8 & 10 (180 s) \\
  La Palma (Canary Islands) & 2m Liverpool & RATCam & $i$ 
        & Feb-May 2007 & 52 & 5 (100 s) \\
  \hline
 \end{tabular}
\end{minipage}
\end{table*}

Optical follow-up of FBQ 0951+2635 has been done in the current decade. This includes 
early imaging and monitoring with the {\it HST} \citep{b9} and the Nordic Optical 
Telescope \citep[{\it NOT},][]{b8}, as well as imaging from the Sloan Digital Sky 
Survey \citep[SDSS,][]{b1}. The 2.5-year monitoring campaign at the {\it NOT} focused on 
the $R$ passband. These $R$-band frames allowed \citet{b8} and \citet{b11} to study the 
time delay between components and early extrinsic variability. The time delay between A 
and B is of about two weeks \citep{b8}, and there is clear evidence for $R$-band extrinsic 
variations in the 1999$-$2001 period \citep{b11}. \citet{b11} showed a possible gradient 
of about 0.1 mmag day$^{-1}$, as well as a possible 50-mmag event on a timescale of several 
months. These extrinsic fluctuations (which are not originated in the source quasar) were 
attributed to microlensing by collapsed objects within the lensing galaxy \citep[e.g.,][and 
references therein]{b13}.

However, the information obtained in the first years of monitoring with the {\it NOT} 
(1999$-$2001) does not permit to decide on the true microlensing variability, so 
additional imaging is required to address this issue. For example, in Fig. 2 (middle and 
bottom panels) of \citet{b11}, one sees that the microlensing gradients can account for 
basically all variability without the need of introducing additional short-term ($\sim$ 
months) microlensing events (see the distributions of points around the linear fits). 
Alternatively, in the same panels of Fig. 2 of \citet{b11}, sets of short-term 
microlensing events can also explain the observed variations. Thus, it is unclear what 
kind of microlensing fluctuations occur in FBQ 0951+2635: short-timescale events or 
gradients lasting years (tracing a long-timescale event)?. Although both kinds of 
fluctuations may be present, one reasonably expects the presence of a dominant kind 
accounting for most of the observed microlensing variability. In this paper, we try
to identify the dominant microlensing variations.

In Sect.~2 we analyse new $R$-band images taken at the Maidanak Observatory in the 
2001$-$2006 period. The previous {\it NOT} results and the Maidanak analysis permit us to 
check the kind of microlensing fluctuations in the red part of the optical continuum (in 
the observer frame). In Sect.~3 we study FBQ 0951+2635 by using a redder optical filter 
($i$ Sloan passband). The $i$-band frames correspond to a new monitoring campaign with the 
2m Liverpool Robotic Telescope ({\it LRT}) during the first semester of 2007. To support
our main results, we also use a few complementary frames in public and private archives. The 
conclusions are presented in Sect.~4.  

\section{Light curve and flux ratio in the $R$ band}

FBQ 0951+2635 is part of a compact lens system, since the two quasar components are 
separated by 1\farcs1 \citep{b12} and the very faint lensing galaxy is 0\farcs2 away from 
the faintest component B \citep{b8}. Fortunately, the lensing galaxy remains too faint 
to be detected with a standard $R$ filter \citep{b8}, and this is an advantage when doing 
photometry. Thus, the system can be described by two stellar-like objects. There are two 
field stars near the lensed quasar: the bright star S1 and the faint star S3 (see 
Fig.~\ref{q951R}). The FWHM of the seeing disc is measured on the S1 star, which is also 
used to estimate the point spread function (PSF) of the stellar-like sources. This 
estimation allows us to obtain PSF fitting photometry for the double quasar.

A single-epoch magnitude difference ($m_B - m_A$) could not represent the magnitude 
difference at the same emission time ($\Delta m_{BA}$), because one should take into 
account the 16-day time delay between components \citep{b8}. However, the typical 
variability of FBQ 0951+2635 on a timescale of $\sim$ 2 weeks gives us the typical 
amplitude of the deviations $\delta = m_B - m_A - \Delta m_{BA}$, so a correction for 
simultaneity can be estimated from variability studies. Throughout most of this paper, we 
derive $\Delta m_{BA}$ values from single-epoch magnitude differences, whose photometric 
uncertainties are properly enlarged to incorporate the simultaneity error $\sigma_{sim}$ 
(i.e., the typical amplitude of $\delta$). We adopt $\sigma_{sim}$ = 0.03 mag, which is 
consistent with the $R$-band root-mean-square (rms) variability of FBQ 0951+2635 over a 
16-day period (see details in subsection 2.1), as well as with the $i$-band variability 
over such timescale (see subsection 3.1). All single-epoch measurements of $\Delta m_{BA}$ 
include the 0.03-mag uncertainty added in quadrature to the photometric errors. The flux 
ratio between components is given by $A/B = 10^{0.4 \times \Delta m_{BA}}$. 

\subsection{Maidanak follow-up observations}

The Maidanak gravitational lens monitoring programme is being conducted by an international 
collaboration of astronomers from Russia, Ukraine, Uzbekistan, and other countries. The 
median FWHM ($\sim$ 0\farcs7) and number of clear nights ($\sim$ 200 nights per year) at 
Mt. Maidanak \citep{b22,b23} permit to obtain high-resolution images of compact lens 
systems \citep[e.g.,][]{b10}, and here we present and analyse $R$-band observations of FBQ 
0951+2635. These Maidanak homogeneous observations from April 2001 to May 2006 are an 
important tool to understand the microlensing variability in the double quasar. 

Our Maidanak monitoring consisted of 190 frames (exposures) in the $R$ band, which were 
taken with the 1.5m AZT-22 Telescope at Mt. Maidanak\footnote{Information on the Maidanak 
Observatory is available from the Website http://www.astrin.uzsci.net/index.html.} 
(Uzbekistan) on 37 different nights (see Table~\ref{MaiLiv}). We used the LN2-cooled CCD 
camera (BroCam) with SITe ST-00A CCD chip. This 2000$\times$800 CCD detector has pixels with 
a physical size of 15 $\mu$m, giving angular scales of 0\farcs135 pixel$^{-1}$ (long-focus 
mode) and 0\farcs268 pixel$^{-1}$ (short-focus mode). The gain and readout noise are 1.2 
e$^{-}$ ADU$^{-1}$ and 5.3 e$^{-}$, respectively. 

\begin{table*}
\centering
\begin{minipage}{150mm}
 \caption{$R$-band photometry of FBQ 0951+2635 from 2001 to 2006.}
 \label{Maiphot}
 \begin{tabular}{@{}ccccccccccccccc@{}}
  \hline
  (1) & (2) & (3) & (4) & (5) & (6) & (7) & (8) & (9) & (10) & (11) & (12) & (13) & (14) & (15) \\
  \hline
10415 & 2 &  904.073 & 150 & 0.94 & 0.03 & 56.0 & 16.884 & 0.001 & 17.948 & 0.015 & 1.063 & 0.033 & 19.490 & 0.013 \\
20304 & 4 & 1237.797 & 180 & 0.96 & 0.05 & 53.0 & 16.990 & 0.008 & 18.056 & 0.030 & 1.065 & 0.048 & 19.462 & 0.012 \\
20311 & 2 & 1244.790 & 240 & 0.99 & 0.19 & 53.0 & 16.987 & 0.003 & 18.098 & 0.039 & 1.111 & 0.052 & 19.469 & 0.004 \\
20314 & 2 & 1247.782 & 180 & 0.89 & 0.10 & 53.0 & 16.995 & 0.005 & 18.104 & 0.012 & 1.108 & 0.031 & 19.448 & 0.035 \\
30427 & 2 & 1656.661 & 180 & 1.06 & 0.12 & 38.0 & 16.995 & 0.000 & 18.271 & 0.009 & 1.276 & 0.031 & 19.458 & 0.030 \\
30521 & 2 & 1680.656 & 180 & 0.89 & 0.08 & 27.0 & 17.028 & 0.000 & 18.210 & 0.014 & 1.182 & 0.033 & 19.406 & 0.056 \\
40116 & 3 & 1920.994 & 180 & 1.04 & 0.06 & 64.0 & 17.055 & 0.011 & 18.266 & 0.026 & 1.212 & 0.048 & 19.464 & 0.035 \\
40122 & 4 & 1926.879 & 180 & 0.91 & 0.11 & 81.0 & 17.062 & 0.003 & 18.306 & 0.007 & 1.244 & 0.031 & 19.455 & 0.011 \\
40209 & 5 & 1944.849 & 180 & 0.97 & 0.05 & 40.0 & 17.072 & 0.002 & 18.296 & 0.014 & 1.224 & 0.033 & 19.474 & 0.014 \\
40213 & 4 & 1948.907 & 210 & 1.00 & 0.13 & 62.5 & 17.047 & 0.005 & 18.304 & 0.009 & 1.257 & 0.033 & 19.462 & 0.008 \\
40219 & 4 & 1954.904 & 180 & 0.99 & 0.19 & 74.5 & 17.021 & 0.004 & 18.346 & 0.009 & 1.325 & 0.031 & 19.460 & 0.012 \\
40222 & 4 & 1957.851 & 180 & 1.01 & 0.17 & 44.5 & 17.038 & 0.002 & 18.328 & 0.018 & 1.289 & 0.036 & 19.467 & 0.026 \\
40226 & 4 & 1961.877 & 180 & 1.05 & 0.10 & 29.0 & 17.059 & 0.006 & 18.365 & 0.013 & 1.306 & 0.035 & 19.483 & 0.022 \\
40227 & 4 & 1962.868 & 180 & 0.91 & 0.09 & 80.5 & 17.036 & 0.003 & 18.330 & 0.009 & 1.294 & 0.031 & 19.439 & 0.007 \\
40301 & 4 & 1965.876 & 180 & 0.93 & 0.20 & 44.0 & 17.046 & 0.002 & 18.333 & 0.013 & 1.287 & 0.033 & 19.455 & 0.013 \\
40409 & 3 & 2004.748 & 180 & 1.13 & 0.11 & 63.3 & 17.092 & 0.008 & 18.207 & 0.017 & 1.115 & 0.038 & 19.483 & 0.013 \\
40413 & 4 & 2012.684 & 180 & 0.93 & 0.05 & 69.0 & 17.097 & 0.001 & 18.295 & 0.005 & 1.198 & 0.030 & 19.488 & 0.020 \\
40504 & 4 & 2029.677 & 180 & 0.97 & 0.05 & 24.5 & 17.100 & 0.006 & 18.327 & 0.011 & 1.227 & 0.034 & 19.504 & 0.016 \\
40511 & 4 & 2036.671 & 180 & 0.99 & 0.08 & 63.5 & 17.068 & 0.005 & 18.337 & 0.009 & 1.270 & 0.033 & 19.444 & 0.012 \\
40520 & 3 & 2045.693 & 180 & 1.04 & 0.04 & 60.0 & 17.078 & 0.004 & 18.302 & 0.009 & 1.225 & 0.032 & 19.448 & 0.007 \\
41204 & 6 & 2243.062 & 180 & 0.90 & 0.09 & 24.3 & 17.208 & 0.005 & 18.446 & 0.010 & 1.239 & 0.031 & 19.480 & 0.029 \\
41205 & 8 & 2243.983 & 180 & 0.89 & 0.14 & 22.8 & 17.217 & 0.004 & 18.448 & 0.012 & 1.231 & 0.032 & 19.479 & 0.012 \\
50208 & 6 & 2309.834 & 180 & 0.88 & 0.06 & 62.0 & 17.241 & 0.003 & 18.469 & 0.007 & 1.228 & 0.031 & 19.473 & 0.008 \\
50214 & 6 & 2315.840 & 180 & 0.97 & 0.05 & 62.3 & 17.226 & 0.003 & 18.430 & 0.012 & 1.205 & 0.033 & 19.475 & 0.013 \\
50306 & 6 & 2335.815 & 180 & 0.97 & 0.03 & 50.0 & 17.251 & 0.004 & 18.481 & 0.006 & 1.231 & 0.031 & 19.478 & 0.016 \\
50414 & 4 & 2374.726 & 180 & 0.90 & 0.04 & 47.5 & 17.236 & 0.006 & 18.459 & 0.007 & 1.224 & 0.032 & 19.430 & 0.006 \\
51126 & 2 & 2601.018 & 300 & 1.09 & 0.03 & 65.0 & 17.339 & 0.014 & 18.488 & 0.015 & 1.150 & 0.041 & 19.489 & 0.017 \\
51128 & 2 & 2602.974 & 300 & 0.92 & 0.09 & 79.0 & 17.322 & 0.003 & 18.569 & 0.019 & 1.247 & 0.034 & 19.420 & 0.010 \\
51201 & 2 & 2606.027 & 300 & 0.90 & 0.10 & 61.0 & 17.337 & 0.002 & 18.576 & 0.002 & 1.239 & 0.030 & 19.438 & 0.003 \\
60402 & 10 & 2727.786 & 180 & 0.83 & 0.15 & 80.2 & 17.372 & 0.002 & 18.649 & 0.005 & 1.278 & 0.030 & 19.377 & 0.006 \\
60411 & 10 & 2736.793 & 180 & 1.11 & 0.15 & 28.4 & 17.511 & 0.017 & 18.457 & 0.036 & 0.946 & 0.060 & 19.481 & 0.017 \\
60413 & 10 & 2738.767 & 180 & 1.06 & 0.08 & 29.0 & 17.433 & 0.006 & 18.588 & 0.024 & 1.155 & 0.042 & 19.423 & 0.024 \\
60417 & 10 & 2742.703 & 180 & 1.05 & 0.13 & 82.0 & 17.428 & 0.003 & 18.610 & 0.007 & 1.182 & 0.031 & 19.401 & 0.005 \\
60509 & 10 & 2764.678 & 180 & 1.04 & 0.17 & 38.6 & 17.451 & 0.003 & 18.637 & 0.011 & 1.185 & 0.032 & 19.429 & 0.011 \\
60511 & 10 & 2766.681 & 180 & 1.17 & 0.10 & 38.2 & 17.529 & 0.019 & 18.422 & 0.043 & 0.893 & 0.069 & 19.437 & 0.014 \\
60514 & 10 & 2769.706 & 180 & 0.97 & 0.16 & 26.2 & 17.483 & 0.009 & 18.682 & 0.012 & 1.198 & 0.032 & 19.453 & 0.024 \\
60517 & 10 & 2772.664 & 180 & 1.14 & 0.09 & 56.6 & 17.472 & 0.004 & 18.528 & 0.011 & 1.055 & 0.033 & 19.416 & 0.011 \\
  \hline
 \end{tabular}

 \medskip
 (1) civil date (ymmdd), (2) number of frames, (3) JD$-$2451100, (4) exposure time per 
frame (s), (5) FWHM of the seeing disc (\arcsec), (6) ellipticity of the seeing disc, (7) 
SNR of an 18 mag star (within a 2\arcsec\ aperture radius), (8) flux of A (mag), (9) 
error in flux of A (mag), (10) flux of B (mag), (11) error in flux of B (mag), (12) $\Delta 
m_{BA}$ (mag), (13) error in $\Delta m_{BA}$ (mag), (14) flux of S3 (mag), and (15) error 
in flux of S3 (mag). 
\end{minipage}
\end{table*}

\begin{figure}
\centering
\includegraphics[angle=0,width=50mm]{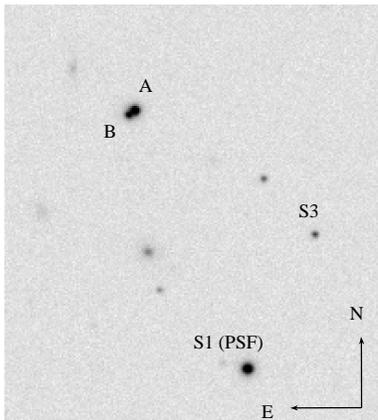}
\caption{Central region of a Maidanak $R$-band frame of FBQ 0951+2635. This frame was taken 
under very good seeing conditions (FWHM = 0\farcs75) on January 22, 2004. The exposure 
time was 180 s, and we show the logarithm of counts in each pixel.}
\label{q951R}
\end{figure}

The pre-processing of each frame consists of bias subtraction, overscan trimming, flat 
fielding, and cosmic rays cleaning. For each observation night, we have two or more 
individual frames obtained under good seeing conditions: $\langle {\rm FWHM} \rangle \sim$ 
1\arcsec. For example, in Fig.~\ref{q951R} we display the central region 
(1\arcmin$\times$1\arcmin) of an 180-s exposure in subarcsecond seeing conditions. This 
$R$-band image (logarithm of counts over the FBQ 0951+2635 field) includes the S1 star 
($\sim$ 16.6 mag), the two components A and B ($\sim$ 17.1 mag and $\sim$ 18.3 mag, 
respectively), and the S3 star ($\sim$ 19.5 mag). Moreover, the average signal-to-noise 
ratio (aperture radius of 2\arcsec, i.e., $2 \times$ FWHM) of an 18 mag star is $\langle 
{\rm SNR} \rangle \sim$ 50. Thus, most of the nightly FWHM values are less than the 
separation of the double quasar and there is significant signal associated with the 
faintest quasar component.

\begin{figure}
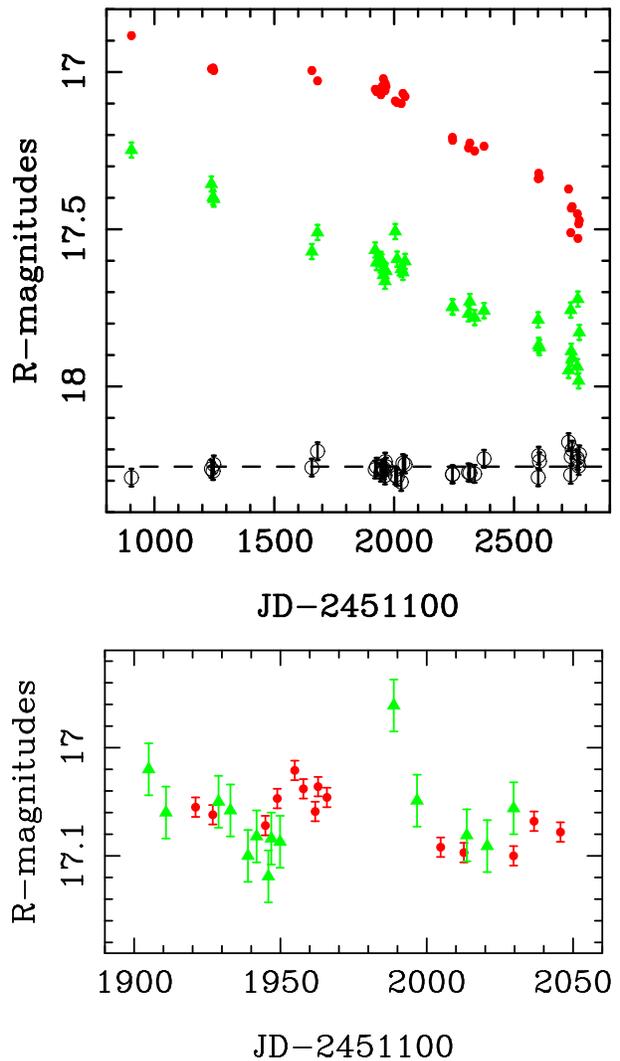

\includegraphics[angle=-90,width=80mm]{f2t.eps}
\vskip 10pt
\includegraphics[angle=-90,width=80mm]{f2b.eps}
\caption{Maidanak $R$-band light curves of FBQ 0951+2635. Top: A (filled circles), B $-$ 
0.7 mag (filled triangles), and S3 $-$ 1.2 mag (open circles). Bottom: comparison between 
the A fluxes in the first semester of 2004 (filled circles) and the time- and 
magnitude-shifted brightness record of B in that semester (filled triangles). See main text 
for details.}
\label{AZTmon}
\end{figure}

PSF fitting photometry leads to instrumental fluxes of components and field stars. These
fluxes are calculated using the instrumental photometry pipeline by \citet{b18}. We also 
obtain calibrated light curves (in mag) of A, B and S3, as well as single-epoch magnitude 
differences $m_B - m_A$. The individual values are then combined to form nightly means and 
standard deviations of means (i.e., standard errors). The standard errors of $m_B - m_A$ 
are properly enlarged because we are interested in magnitude differences at the same 
emission time (see explanation above). The fluxes of A, B and S3 on April 15, 2001 are also 
compared to the corresponding {\it NOT} fluxes at close dates \citep{b8}. We do not find 
offsets in the fluxes of A and S3, but the flux of B is slightly higher (60 mmag) than the 
{\it NOT} level around April 15, 2001. This offset is very probably due to a small 
contamination by the lens galaxy light (which is absent in the {\it NOT} fluxes), so a 
60-mmag correction is taken into account when obtaining final fluxes of B. Each of the four 
combined curves ($m_A$, $m_B$, $m_{S3}$ and $\Delta m_{BA}$) incorporates 37 data points. 
These data points and other quantities of interest are presented in Table~\ref{Maiphot}. 

Although the standard errors may be good estimators of the nightly photometric errors in 
$m_A$, $m_B$ and $m_{S3}$, changes in the colour coefficient and/or the possible 
inhomogeneous response over the camera area could produce a substantial amount of 
additional noise throughout the several years of monitoring \citep[e.g.,][]{b18}. We check 
this possibility by comparing the average standard error of $m_{S3}$ and the standard 
deviation of $m_{S3}$ over the whole monitoring period. There is a bias factor of about 
$\sqrt{3}$, so we multiply the average standard errors of $m_A$, $m_B$ and $m_{S3}$ by 
$\sqrt{3}$ to derive typical photometric errors. These final uncertainties are 9 mmag (A), 
24 mmag (B) and 28 mmag (S3). The top panel in Fig.~\ref{AZTmon} shows the final brightness 
records of the double quasar and the S3 star. Filled circles, filled triangles and open 
circles trace the behaviours of A, B (shifted by $-$0.7 mag) and S3 (shifted by $-$1.2 mag), 
respectively. The almost parallel fading by $\sim$ 0.6 mag of the two components indicates 
the presence of long-timescale intrinsic variability (extrinsic variations are discussed 
below). Around day 2750 (JD$-$2451100), one can see an important scatter in the light 
curves of A and B. This scatter was produced when the camera (BroCam) reached the end of 
its lifetime.  

In general, sampling is unsuitable when doing analysis of short-timescale (intra-year) 
variability. However, this kind of study is possible in 2004 (see Table~\ref{MaiLiv}), i.e., 
around day 2000 in Fig.~\ref{AZTmon} (top panel). In the bottom panel of Fig.~\ref{AZTmon}, 
while the filled circles describe the time evolution of the flux of A, the filled triangles 
represent the time- and magnitude-shifted light curve of B. The fluxes of B are advanced in 
16 days \citep[time delay;][]{b8} and increased by 1.246 mag (average magnitude difference 
between the time-shifted record of B and the light curve of A; although bins with semi-size
of 4 days are used, the average magnitude difference is within the 1.24--1.25 mag interval 
for any bin semi-size below 6 days). We show that there is good agreement between both 
brightness records, i.e., the two records are consistent with each other in the overlap 
periods. For this reason, we can state that short-timescale microlensing is absent or 
elusive. The time-delay-corrected flux ratio ($R$-band data from day 1900 to day 2050) is 
$A/B$ = 3.15 $\pm$ 0.05 (1$\sigma$).

We also obtain the Maidanak and {\it NOT} $R$-band typical variability of FBQ 0951+2635 
at time lags $\Delta t \leq$ 100 days, with special emphasis on $\Delta t$ = 16 days, i.e., 
the time delay between components. The typical magnitude variation on this timescale is 
directly related to the simultaneity error $\sigma_{sim}$ that we introduce in the second 
paragraph of section 2. The rms variability $\langle (\Delta s)^2 \rangle^{1/2}$ at lag 
$\Delta t$ is given by
\begin{equation}
\langle (\Delta s)^2 \rangle^{1/2} = \{ (1/N) \sum_{i,j} [(m_j - m_i)^2 - \sigma_i^2 - 
\sigma_j^2] \}^{1/2}  ,
\end{equation}
where the sum only includes the ($i$,$j$) pairs verifying that $t_j - t_i \sim \Delta t$ 
(the number of such pairs is $N$). To derive the rms values from Eq. (1), we consider the
magnitude variations in both components A and B (Maidanak fluctuations from day 2700 to
day 2800 are doubtful, and thus, this Maidanak observation period is not taken into account 
in the analysis; see above). Fig.~\ref{sf} shows the structure functions obtained from the 
Maidanak (filled circles) and {\it NOT} (filled squares) light curves, using independent 
4-day bins. Two vertical dashed lines define the bin of interest (around the lag of 16 days). 
At lags $\Delta t <$ 60 days, the quasar was somewhat less active over days 50$-$1000 ({\it 
NOT} data). The horizontal dashed line corresponds the 0.03 mag level, which roughly
coincides with the Maidanak rms variability at the lag centered on 16 days. These results
explain why we take $\sigma_{sim}$ = 0.03 mag, at least at red wavelengths. We remark that 
the bottom panel of Fig.~\ref{AZTmon} shows a "flare" of $\sim$ 0.1 mag over $\sim$ 10 days, 
between day 1990 and day 2000, and this seems to question our variability analysis in 
Fig.~\ref{sf}. However, taking the error bars into account, the true variation in the flux 
of B could be of about 30 mmag. This is in good agreement with the adopted variability over 
a 16-day period. Moreover, the $\sim$ 0.1-mag jump (considering central values) is only 
defined by two consecutive fluxes of the B component. In other words, it is a very poorly 
sampled variation in the faintest and noisest component, which is most likely caused by 
observational noise. 

The 37 single-epoch measurements of $\Delta m_{BA}$ in Table~\ref{Maiphot} are grouped in 7 
different time intervals (see the seven periods in Table~\ref{MaiLiv} and the corresponding
days in Table~\ref{Maiphot}). We compute the weighted average and its uncertainty for each 
interval. These averages are then translated into 7 $R$-band flux ratio measurements with an 
1$-$3\% accuracy. The new flux ratio values are depicted in Fig.~\ref{fratio} (filled 
circles). For example, the flux ratio around day 2000 (3.17 $\pm$ 0.03; 1$\sigma$) is 
consistent with the result in two paragraphs above, which takes the time delay correction 
into account. Around day 2750, we obtain two estimates of $A/B$ due to the fact that the 
photometric data show a large scatter (see above). Firstly, we use all available nights 
(last open circle in Fig.~\ref{fratio}). Secondly, only high-quality nights, i.e., FWHM $<$ 
1\farcs1 and SNR $>$ 50, are considered (filled circle above the last open circle in 
Fig.~\ref{fratio}). This second estimation seems more reliable, so the first one is assumed 
to be biased. We also compare the {\it NOT} flux ratio at an intermediate epoch and the 
corresponding Maidanak ratio from the former ST-7 CCD camera (which was operating in that 
epoch). The Maidanak/ST-7 1$\sigma$ measurement, 2.58 $\pm$ 0.05 (first open circle at day 
484), is in excellent agreement with the {\it NOT} gradient and limits (dashed lines; see 
the first paragraph in subsection 2.3).

\subsection{Complementary frames}

A few complementary frames are used to check the reliability of the $A/B$ values from the
Maidanak/BroCam homogeneous monitoring. The SDSS archive\footnote{See the SDSS Data Release 
6 (DR6), which incorporates the DR6 Catalogue Archive Server site at 
http://cas.sdss.org/astrodr6/en/.} \citep{b1} contains a frame of FBQ 0951+2635 in the $r$ 
band. This frame was taken under very good seeing conditions (FWHM = 0\farcs83) on December 
11, 2004. From PSF fitting photometry, we infer $m_B - m_A$ = 1.22 $\pm$ 0.01 mag (1$\sigma$), 
which translates into $\Delta m_{BA}$ = 1.22 $\pm$ 0.03 mag and $A/B$ = 3.08 $\pm$ 0.08. The
$r$-band flux of B is not corrected by a possible contamination by the lens galaxy light 
(the $r$-band correction will be significantly smaller than the 60-mmag offset in the $R$ 
band, since the $r$ Sloan filter does not transmit light at wavelengths of 700--900 nm), and 
the $r$-band magnitude difference is taken as the magnitude difference in the $R$ band (this 
is a reasonable assumption because the $r - R$ colours of both components are similar). We 
note the agreement between the SDSS flux ratio at day 2250 (first filled triangle in 
Fig.~\ref{fratio}), and the Maidanak/BroCam value of $A/B$ around day 2300 (filled circle next 
to the first filled triangle in Fig.~\ref{fratio}). 

Maidanak observations in 2007 were done with the Spectral Instruments (SI) 600 Series CCD 
camera. It is a 4096$\times$4096, 15-$\mu$m pixel CCD. The rectangular pixels have angular 
scales of 0\farcs303 (along the horizontal axis) and 0\farcs258 (along the vertical axis). 
The camera gain is 1.45 e$^{-}$ ADU$^{-1}$ and the readout noise is 4.7 e$^{-}$. These
Maidanak/SI 600 Series frames belong to a private archive that is managed by several teams.
To test the time evolution of $A/B$ in recent years, we use the three $R$-band frames that 
were taken on April 13, 2007. After applying a 60-mmag correction to the fluxes of B, the
flux ratio at day 3104 is $A/B$ = 3.17 $\pm$ 0.10 (last filled triangle in 
Fig.~\ref{fratio}).  
 
\subsection{Global perspective on flux ratio evolution}

\begin{figure}
\includegraphics[angle=-90,width=80mm]{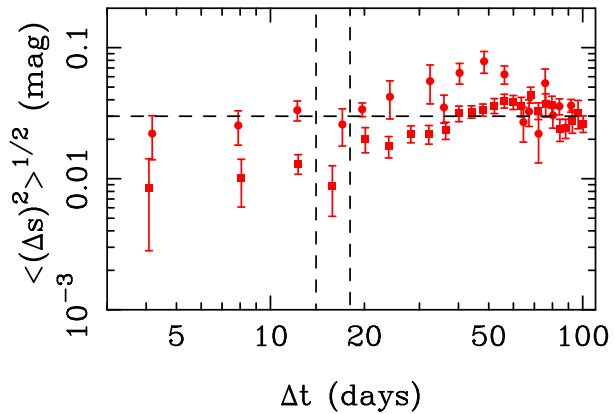}
\caption{$R$-band rms variability at different time lags. The Maidanak (filled circles) 
and {\it NOT} (filled squares) trends are analysed in a separate way. We use 4-day 
independent bins centered on lags of 4, 8, 12, 16, 20,... days. The bin of interest (around
the lag of 16 days) is highlighted from two vertical dashed lines, and the horizontal dashed
line represents the 0.03 mag level.}
\label{sf}
\end{figure}

\begin{figure}
\includegraphics[angle=-90,width=80mm]{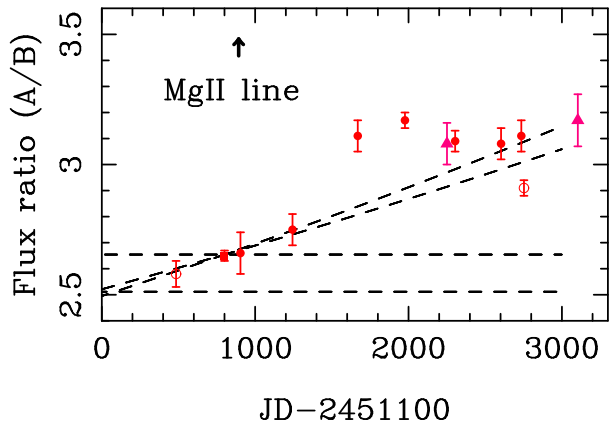}
\caption{Time evolution of the optical continuum (red wavelengths) flux ratio of FBQ 
0951+2635. We show a previous measurement from {\it NOT} $R$-band frames (filled square 
at day 800), as well as new Maidanak/BroCam $R$-band estimates (filled circles) and two
additional results from SDSS and Maidanak/SI 600 Series frames (filled triangles). We also 
show the lower limit of the \hbox{Mg\,{\sc ii}} flux ratio (base of the vertical arrow). 
The open circles and dashed lines are explained in the main text.}
\label{fratio}
\end{figure}

\citet{b11} suggested two possible $R$-band microlensing gradients over days 50$-$1000 
\citep[see middle and bottom panels in Fig. 2 of][]{b11}. However, the second solution 
(0.077 $\pm$ 0.007 mmag day$^{-1}$) seems more consistent with the analysis by \citet{b8}, 
so we only consider this early microlensing slope. \citet{b8} found a flux ratio $A/B$ = 
2.65 $\pm$ 0.02 ($\Delta m_{BA}$ = 1.06 $\pm$ 0.01 mag) when they focussed on the last 
part of the {\it NOT} brightness records, i.e., around day 800 (see the filled square in 
Fig.~\ref{fratio}). This is fully consistent with the second linear fit by \citet{b11} 
(see bottom panel in Fig. 2 of that work), which indicates $\Delta m_{BA}$ = 1.06 at day 
800. Alternatively, the extrinsic variability could be due to two consecutive 
short-term microlensing events before day 700. While the first event was fitted to a 
Gaussian, it is also apparent the possible existence of a second event (see data point 
distribution between days 400 and 600). From day 700 to day 1000, the observations are 
consistent with constant microlensing because almost all error bars cross the zero level. 
If the two rapid events with similar amplitude are true features, no gradient is required 
to account for the observations. The two alternatives and their extrapolations to more 
recent epochs are depicted in Fig.~\ref{fratio}. The inclined dashed lines represent the 
gradient of 0.077 $\pm$ 0.007 mmag day$^{-1}$ (absence of short-term events). The 
horizontal dashed lines are the lower and upper limits associated with the second scenario: 
absence of a gradient, but presence of short-timescale fluctuations with amplitude of 
about 60 mmag (flaring behaviour of B). 

As we can see in Fig.~\ref{fratio}, the Maidanak measurements from day 2600 to day 3100, 
i.e., the last two filled circles and the last filled triangle, are consistent with the 
first microlensing scenario. However, the most recent behaviour of $A/B$ does not imply 
the presence of a long-term gradient lasting $\sim$ 10 years. The whole set of results 
within the $\sim$ 3000-day monitoring period suggests the existence of a long-timescale 
microlensing fluctuation, which had a bump between day 1200 and day 2200, and has remained 
almost flat during the last years. The bump consists of a significant increase (in $A/B$) 
of about 15\% followed by a shallow decrease. As the flux ratio is measured accurately, 
the signal-to-noise ratio for the prominent increase is $\sim$ 3$-$5. When a source 
crosses a microlensing magnification pattern, the associated light curve may have a complex 
structure including different gradients at different epochs \citep[e.g.,][]{b13}. We also 
point out that our new flux ratio estimates clearly disagree with the second scenario (see 
horizontal dashed lines in Fig.~\ref{fratio}), which is based on the exclusive production 
of short-term ($\sim$ months) fluctuations.

\section{Follow-up in the \lowercase{$i$} Sloan passband}

\subsection{LQLM campaign}

\begin{table*}
\centering
\begin{minipage}{150mm}
 \caption{$i$-band photometry of FBQ 0951+2635 in 2007.}
 \label{Livphot}
 \begin{tabular}{@{}ccccccccc@{}}
  \hline
  (1) & (2) & (3) & (4) & (5) & (6) & (7) & (8) & (9) \\
  \hline
70207 & 5 & 3039.691 & 1.07 & 0.17 & 80 & 17.507 & 18.556 & 18.508 \\ 
70208 & 5 & 3040.391 & 1.27 & 0.14 & 78 & 17.413 & 18.622 & 18.523 \\ 
70209 & 5 & 3041.378 & 1.41 & 0.07 & 73 & 17.456 & 18.563 & 18.519 \\ 
70210 & 5 & 3042.411 & 1.32 & 0.32 & 96 & 17.479 & 18.578 & 18.521 \\ 
70211 & 5 & 3043.414 & 1.28 & 0.00 & 74 & 17.455 & 18.570 & 18.502 \\ 
70212 & 4 & 3044.380 & 1.15 & 0.04 & 80 & 17.458 & 18.657 & 18.527 \\ 
70218 & 5 & 3050.370 & 1.4  & 0.31 & 96 & 17.465 & 18.593 & 18.533 \\ 
70220 & 5 & 3052.371 & 1.21 & 0.09 & 79 & 17.469 & 18.566 & 18.531 \\ 
70221 & 5 & 3053.396 & 1.25 & 0.22 & 85 & 17.474 & 18.498 & 18.525 \\ 
70222 & 4 & 3054.389 & 1.3  & 0.02 & 68 & 17.506 & 18.571 & 18.520 \\
70223 & 5 & 3055.388 & 1.05 & 0.34 & 86 & 17.416 & 18.565 & 18.516 \\ 
70404 & 5 & 3095.492 & 1.26 & 0.05 & 58 & 17.468 & 18.529 & 18.500 \\
70406 & 4 & 3097.517 & 0.97 & 0.07 & 76 & 17.505 & 18.659 & 18.471 \\ 
70413 & 3 & 3104.478 & 1.09 & 0.06 & 94 & 17.462 & 18.529 & 18.493 \\ 
70414 & 4 & 3105.494 & 1.18 & 0.05 & 99 & 17.470 & 18.560 & 18.507 \\ 
70415 & 5 & 3106.461 & 1.37 & 0.13 & 78 & 17.459 & 18.529 & 18.517 \\ 
70421 & 5 & 3112.430 & 1.22 & 0.1  & 87 & 17.452 & 18.512 & 18.497 \\ 
70422 & 4 & 3113.411 & 1.22 & 0.04 & 58 & 17.406 & 18.567 & 18.509 \\
70504 & 4 & 3125.402 & 1.19 & 0.06 & 81 & 17.430 & 18.570 & 18.510 \\
70507 & 4 & 3128.429 & 1.29 & 0.03 & 68 & 17.416 & 18.524 & 18.492 \\ 
70512 & 4 & 3133.428 & 1.32 & 0.11 & 62 & 17.422 & 18.580 & 18.498 \\ 
70528 & 5 & 3149.463 & 1.25 & 0.04 & 54 & 17.436 & 18.528 & 18.471 \\
  \hline
 \end{tabular}

 \medskip
 (1) civil date (ymmdd), (2) number of 100-s exposures, (3) JD$-$2451100, (4) FWHM of 
the seeing disc (\arcsec), (5) ellipticity of the seeing disc, (6) SNR of the S3 star 
(within a $2 \times$ FWHM aperture radius), (7) flux of A (mag), (8) flux of B (mag), 
and (9) flux of S3 (mag). 
\end{minipage}
\end{table*}

We included FBQ 0951+2635 as a key target in our Liverpool Quasar Lens Monitoring (LQLM) 
programme \citep{b7,b18}. The optical frames were taken with the {\it 
LRT}\footnote{Information on the Liverpool Robotic Telescope is available from the Website 
http://telescope.livjm.ac.uk/.} between February 6, 2007, and May 31, 2007. Each 
observation night consisted of five exposures of 100 sec in the $i$ band, using a dither 
cross pattern (see observations summary in Table~\ref{MaiLiv}). The {\it LRT} 
pre-processing pipeline included bias subtraction, trimming of the overscan regions, and 
flat fielding. These pre-processed frames are publicly available on the Lens Image Archive 
of the German Astrophysical Virtual Observatory\footnote{See the Web site 
http://vo.uni-hd.de/lensdemo/view/q/form.}. We initially remove all frames that either are 
characterised by an anomalous image formation or have large seeing discs (FWHM $>$ 
2\arcsec\ in the frame headers). Later we carry out cosmic rays cleaning and defringing. 
In the last step of the whole pre-processing procedure, all frames in each night ($\leq$ 5) 
are combined, i.e., they are aligned and then averaged. 

We create a stacked frame (consisting of the combination of the best exposures) to better 
detect the lens galaxy and to subtract its light in the combined frames. This represents a 
total exposure time of 4.4 hours. The stacked image is characterised by FWHM = 1\farcs17 
and a very high signal-to-noise ratio (SNR = 413), where FWHM and SNR (aperture radius of 
$2 \times$ FWHM) are measured on the S1 and S3 stars, respectively. In the $i$ band, S3 has 
a magnitude similar to that of the faintest quasar component B. Unfortunately, neither our 
best combined frames in terms of FWHM and SNR, nor our deep stacked frame lead to detection 
of the lens galaxy. Moreover, we obtain meaningless results by using constraints on the 
relative position and brightness profile of the galaxy \citep[taken from][]{b8}. The 2D 
signal from the very faint galaxy seems to be strongly affected by the photon noise of the
quasar components and background, so we cannot resolve that signal in our frames. Thus, 
our instrumental photometry pipeline \citep{b18} only fits the instrumental fluxes of both 
components, which are described by stellar-like objects (PSF fitting photometry, using S1 
as PSF star). Once the instrumental photometry is done, we obtain calibrated and corrected 
brightness records of A, B and field stars. The transformation (calibration-correction) 
pipeline incorporates zero-point, colour and linear inhomogeneity terms, and the final 
magnitudes are given in the SDSS photometric system \citep{b18}. This transformation 
software is exclusively applied to the combined frames with FWHM $<$ 1\farcs5 and SNR $>$ 
50, i.e., on 22 out of 52 observation nights (see Table~\ref{Livphot}).

\begin{figure}
\includegraphics[angle=0,width=70mm]{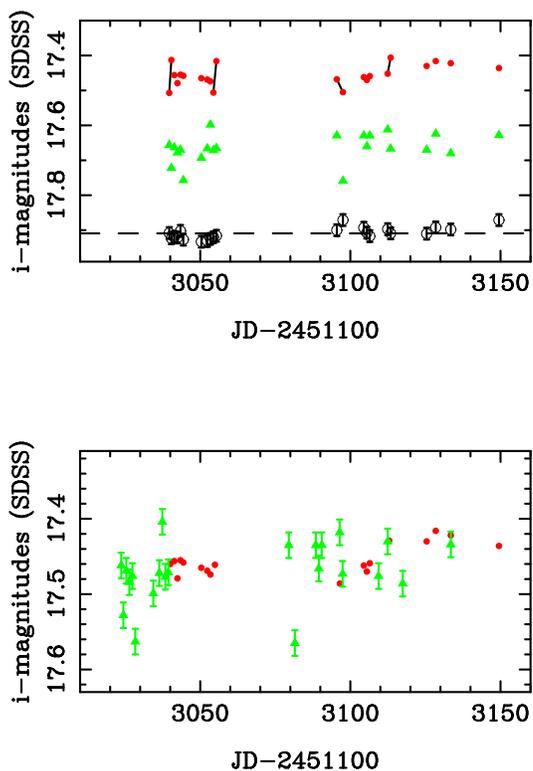}
\caption{LQLM $i$-band light curves of FBQ 0951+2635. Top: A (filled circles), B $-$ 0.9 
mag (filled triangles), and S3 $-$ 0.6 mag (open circles). Bottom: comparison between the
A master record (filled circles) and the time- and magnitude-shifted light curve of B 
(filled triangles). See main text for details.}
\label{LTrmon}
\end{figure}

The top panel in Fig.~\ref{LTrmon} displays the LQLM light curves of the lensed quasar (A
and B components) and the S3 star. Filled circles, filled triangles and open circles trace 
the behaviours of A, B (shifted by $-$0.9 mag) and S3 (shifted by $-$0.6 mag), 
respectively. There must be some diffuse contamination by the lens galaxy light, so the B 
fluxes in Table 3 and Fig.~\ref{LTrmon} are greater than the true ones. From the standard 
deviation of $m_{S3}$ over the whole monitoring period, we infer a typical error in the S3 
fluxes of 17 mmag (see open circles and associated error bars). Hence, the typical 
uncertainty in the B fluxes should be larger than 17 mmag, since B is as bright as S3 
($\sim$ 18.5$-$18.6 mag), but it is located within a crowded region. Unfortunately, even 
using frames with $\langle {\rm FWHM} \rangle \sim$ 1\farcs2 and $\langle {\rm SNR} \rangle 
\sim$ 80, we cannot achieve 1$-$2\% level photometry for B (some deviations between 
adjacent nights are relatively large). 

Four pairs of adjacent fluxes of A show significant scatters around the mean values (in 
the top panel in Fig.~\ref{LTrmon}, we draw four lines joining the members of each pair).
For this reason, they are grouped (by computing four mean fluxes at the corresponding
mean epochs) to obtain a master light curve, i.e., an accurate (smooth) trend that 
reliably describes the underlaying short-timescale variability. This master curve is 
depicted in the bottom panel of Fig.~\ref{LTrmon} (filled circles). Before the gap, 
the quasar has a small level of activity. However, after the gap, there is a 60-mmag 
gradient lasting about 30 days. Thus, the $R$-band and $i$-band records exhibit similar 
variability levels on a timescale of $\sim$ 2 weeks (see Fig.~\ref{sf}). To gain 
perspective on the nature of the observed variability in the $i$ band (A component), the B 
light curve is shifted in time \citep[using the 16-day time delay by][]{b8}, and then is 
compared to the master light curve of A. We derive an average magnitude difference 
of 1.094 mag between the time-shifted record of B and the master curve of A (using bins 
with semi-size of 2--4 days). The time- and magnitude-shifted light curve of B is also 
included in the bottom panel of Fig.~\ref{LTrmon} (filled triangles). The typical error 
in the S3 fluxes is used as a lower limit for the uncertainty in the photometric 
measurements of B (see above). We find a clear agreement between the master curve of A 
and the adjacent fluxes of B, when these last fluxes are properly shifted in time and 
magnitude. Thus, there is no evidence of short-timescale extrinsic variability (due to 
microlensing or another phenomenon) in our $i$-band observations.

The new LQLM light curves of FBQ 0951+2635 allow us to measure an accurate flux ratio 
(corrected by the time delay) in the $i$ band: $A/B_*$ = 2.74 $\pm$ 0.02 (1$\sigma$).
We do not measure the $i$-band flux ratio $A/B$ around day 3100 but the contaminated ratio 
$A/B_*$, where $B_* = B + G$ and $G$ is a contribution due to the lensing galaxy.

\subsection{SDSS frame and recent evolution of $A/B$}

The high-quality SDSS$^2$ $i$-band frame of FBQ 0951+2635 at day 2250 (FWHM = 0\farcs75 
and SNR = 99) leads to a flux ratio $A/B_*$ = 2.8 $\pm$ 0.1 (1$\sigma$), in good 
agreement with the LQLM estimation. We note that $B_* = B + G$ (see above), and the 
uncertainty in the SDSS flux ratio incorporates both photometric and simultaneity errors. 
This last error is associated with the use of fluxes at the same time of observation, 
i.e., without time delay correction (see above). There is no evidence of an 
appreciable evolution in $A/B_*$ over days 2250$-$3100. This $i$-band result agrees with 
the Maidanak-SDSS recent trend of $A/B$ in the $R$ band (see filled circles and triangles 
from day 2250 to day 3100 in Fig.~\ref{fratio}). Comparing the flux ratios in both optical
filters, we may derive that the lens galaxy light in the $i$ band produces a plausible 
contamination of the B component in such filter ($\sim$ 130 mmag).   

\section{Conclusions}

A previous analysis \citep{b11} showed the existence of extrinsic variability of the 
observer-frame optical continuum in the gravitationally lensed double quasar FBQ 
0951+2635. The two quasar components (A and B) cross two different regions of the lensing 
galaxy, so distributions of collapsed objects could affect one (or both) of the light 
curves and then produce extrinsic variations. This microlensing hypothesis is more 
plausible for the B component because it crosses the central region of the galaxy 
\citep[see][]{b8}. \citet{b11} reported on two possible microlensing variabilities at 
red wavelengths in the 1999$-$2001 period: short-timescale events (B flares having a 
duration of months) or a gradient lasting a few years. Both alternatives can account for 
the 1999$-$2001 $R$-band observations, and this paper sheds light on the dominant kind of 
microlensing fluctuations occurring in FBQ 0951+2635.  
  
We analyse new Maidanak $R$-band images taken in the 2001$-$2006 period (a 6-year 
homogeneous monitoring), and a few complementary frames in the $rR$ bands taken in 2004 and 
2007 (SDSS and Maidanak/SI 600 Series archives). The Maidanak-SDSS flux ratios ($A/B$) in 
the red part of the optical continuum are inconsistent with the absence of long-timescale 
($\sim$ years) gradients and the continuous production of B flares lasting a few months 
(short-timescale microlensing events). If this last scenario were true, all data points in 
Fig.~\ref{fratio} would be distributed between the two horizontal dashed lines. The whole 
set of flux ratios (from 1999 to 2007) favour the existence of a long-timescale microlensing 
fluctuation, so long-timescale gradients seem to be the dominant microlensing variations. 
While $A/B$ shows a bump in the 2003$-$2004 period, it is almost constant from late 2004 to 
the middle of 2007. This last quasi-stationary behaviour is supported by additional data in 
the $i$ band (from new LQLM images taken in 2007 and the SDSS archive frame in that filter). 
\citet{b17} and \citet{b11} also found a long-timescale microlensing variation in the doubly 
imaged quasar SBS 1520+530. Although the existence of rapid flares in the early years of 
monitoring (1999$-$2001) cannot be ruled out, these hypothetical B flares are not produced 
in a continuous way. Short-timescale microlensing is not detected in the Maidanak 
variability study over the first semester of 2004. The LQLM data in the $i$ band also 
indicate the absence of short-timescale events in 2007. Apart from the 1999$-$2007 optical 
data, FBQ 0951+2635 has recently been observed in X-rays \citep{b2}. However, the X-ray flux 
ratio has a large uncertainty and it is not useful for comparison with our estimates in the 
optical continuum. 

If microlensing in the B component is the physical origin of the long-timescale fluctuation, 
this fluctuation represents a progressive demagnification of B followed by a 
quasi-stationary evolution in more recent epochs. The optical continuum flux ratio at red 
wavelengths is always below the \hbox{Mg\,{\sc ii}} flux ratio \citep[][see also 
Fig.~\ref{fratio}]{b8}, which is usually assumed to be weakly affected by microlensing.
Thus, the recent microlensing magnification of B would be relatively weak, but still 
appreciable. The microlensing peak (maximum magnification of B) would have taken place 
before the discovery of the lens system by \citet{b12}. Strictly speaking, the continuum 
flux ratio refers to the flux ratio in the $R$ band, i.e., a spectral region containing 
both the red continuum and the \hbox{Mg\,{\sc ii}} emission line. However, this line only 
contributes a few percent to the broad band fluxes \citep[see Fig. 4 of][]{b12}. It can be
shown that the line-corrected ratio is very similar to the ratio from total fluxes in the
$R$ band \citep[see also][]{b8}. We also note that $A/B$ is not corrected by any 
extinction factor, so this flux ratio might be a biased estimator of the lens (macrolens + 
microlens) magnification ratio. For example, \citet{b4} suggested the relative extinction 
of A by dust in the lensing galaxy. This dusty scenario leads to a true lens magnification 
ratio greater than $A/B$.

We point out that new flux ratios at a large collection of wavelengths should be key tools 
to discuss dust extinction and microlensing in the lensing galaxy 
\citep[e.g.,][]{b14,b4,b5,b6,b20,b16}. Moreover, FBQ 0951+2635 must be imaged in the $rR$ bands 
during the next years (several times per year, with a 2-week separation between consecutive
observations). This modest monitoring programme will lead to draw the future evolution of 
$A/B$, as well as to obtain relevant information on the structure of both the source quasar 
and the intervening galaxy \citep[e.g.,][]{b13,b19,b21}. 

\section*{Acknowledgments}

We thank an anonymous referee for several comments that improved the presentation of our 
results. The Uzbek team thanks Prof. J. Wambsganss and R. Schmidt for helping them in the 
operation of the Maidanak Observatory during the 2003$-$2004 period. The AZT-22 Telescope 
at Mt. Maidanak (Uzbekistan) is owned by the Ulugh Beg Astronomical Institute of the Uzbek 
Academy of Sciences, and operated by an international collaboration. The Liverpool Telescope 
is operated on the island of La Palma by Liverpool John Moores University in the Spanish 
Observatorio del Roque de los Muchachos of the Instituto de Astrofisica de Canarias with 
financial support from the UK Science and Technology Facilities Council. We thank C. Moss 
for guidance in the preparation of the robotic monitoring project with the Liverpool 
Telescope (CL07A04 programme). We also use information taken from the Sloan Digital Sky 
Survey (SDSS) Web site, and we are grateful to the SDSS team for doing that public database. 
This research has been supported by the Spanish Department of Education and Science grant 
AYA2007-67342-C03-02, University of Cantabria funds, the Russian Foundation for Basic 
Research (RFBR) grants No. 06-02-16857 and 09-02-00244, the Taiwan National Research 
Councils grant No. 96-2811-M-008-058, and the Uzbek Academy of Sciences grant No. FA-F2-F058.

\bsp

\label{lastpage}

\end{document}